\documentclass[12pt]{iopart}
\usepackage{graphicx}
\usepackage{cite}
\usepackage{amsfonts}
\usepackage{array}
\usepackage{booktabs}

\begin{document}
\title[Is local scale invariance a generic property of ageing phenomena ?]
      {Is local scale invariance a generic property of\\ ageing phenomena ?}
\author{Haye Hinrichsen}

\address{Fakult\"at f\"ur Physik und Astronomie\\
         Universit\"at W\"urzburg\\ D-97074 W\"urzburg, Germany}

\begin{abstract}
In contrast to recent claims by Enss, Henkel, Picone, and Schollw{\"o}ck [J. Phys. A {\bf 37}, 10479] it is shown by numerical simulations that the autoresponse function of the critical 1+1-dimensional contact process is not in agreement with the predictions of local scale invariance.
\end{abstract}

\submitto{Journal of Statistical Mechanics: Theory and Experiment}
\parskip 2mm 
\vglue 5mm

\noindent
The contact process is a simple lattice model that is often used to describe the spreading of an infectious disease~\cite{Harris,Marro,Hin}. It is defined on a hypercubic lattice whose sites can be either active or inactive. The model evolves in time by random-sequential updates in a way that each active site either spontaneously activates a nearest-neighbor site with rate $\lambda$ or it becomes inactive with rate $1$. Depending on the control parameter $\lambda$ the (infinite) model exhibits a phase transition from a fluctuating active phase into a completely inactive absorbing state. This transition is continuous and belongs to the universality class of directed percolation (DP). 

Recently Enss, Henkel, Picone, and Schollw{\"o}ck studied the 1+1-dimensional contact process as a generic example of ageing phenomena without detailed balance~\cite{Malte1}. In particular, they investigated the autoresponse function
\begin{equation}
R(t,s) = \left.\frac{\delta \langle \phi(\vec{x},t) \rangle}{\delta h(\vec{x},s)}\right|_{h(\vec{x})=0}
\end{equation}
in a \textit{critical} system starting with a fully occupied lattice at $t=0$. The autoresponse function quantifies how strongly a local perturbation at position $\vec{x}$ exerted by an external field $h(\vec{x},s)$  at a given time $s>0$ changes the averaged order parameter $\langle\phi(\vec{x},s)\rangle$ at some later time $t>s$. In the contact process the order parameter is the density of active sites $\langle n_i(t)\rangle$, where $n_i(t)=0,1$ denotes the state of site~$i$ at time~$t$, while its conjugated field corresponds to spontaneous activation of inactive sites. Hence, the autoresponse function may be determined numerically by simulating a critical contact process starting with a fully occupied initial lattice, then activating a random site $i$ at time $s$ (by setting $n_i(s):=1$), and measuring how much this intervention increases the average activity $\langle n_i(t) \rangle$ at the \textit{same} site at a later time $t>s$.

It is generally accepted that in scale-invariant systems the response function can be expressed in terms of the scaling form
\begin{equation}
R(t,s)  \sim s^{-1-a} f_R(t/s)\,,
\end{equation}
where $f_R(\xi)$ is a universal scaling function and the exponent $a$ depends on the scaling dimension of the order parameter and the response field. In the case of the 1+1-dimensional contact process this exponent is given by $a=2\delta-1 \approx -0.681$, where $\delta=\beta/\nu_\parallel$ is the usual decay exponent of DP (for a definition of DP exponents see e.g. Ref.~\cite{Hin}). 

The scaling function $f_R(\xi)$ is usually characterized by certain asymptotic power laws. In the case of the contact process these power laws are given by
\begin{equation}
\label{AsymptoticPowerLaw}
f_R(\xi) \sim 
\left\{
\begin{array}{ll}
\xi^{-\lambda_R/z} & \mbox{ for } \xi \to \infty\\
(\xi-1)^{-2\delta} & \mbox{ for } \xi \to 1
\end{array}
\right.\,,
\end{equation}
where $\lambda_R$ is the so-called autoresponse exponent and $z=\nu_\parallel/\nu_\perp$ denotes the usual dynamical critical exponent of DP. The value of the autoresponse exponent was estimated numerically by $\lambda_R/z=1.76(5)$.\footnote{After submission Gambassi~\cite{Gambassi} proposed that the autoresponse exponent of critical DP may be expressed in terms of standard DP exponents as $\lambda_R/z=1+\delta+d/z$. This scaling relation would give the value $1.7921$ in one dimension, consistent with the numerical estimate used in the present work.} 

\noindent
It is important to note that in the \textit{critical} contact process the two limits of
\begin{quote}
\begin{itemize}
\item[(a)] first taking $t,s \to \infty$ with $\xi=t/s$ fixed and then sending $\xi \to 1$, and
\item[(b)] sending $t,s \to \infty$ while keeping the difference $t-s$ fixed
\end{itemize}
\end{quote}
give identical results. In the context of ageing phenomena this seems to be surprising since in most cases studied so far these limits are known to be different~\cite{Zippold}. However, if these limits were different in the present case, it would require the existence of a crossover from the quasi-stationary to the ageing regime at some typical time scale $t_c$. This crossover time $t_c$ may be either constant (much larger than the microscopic time scale) or it may grow slowly with $s$ such that $t_c/s \to 0$. Obviously, in a scale-free system such as the contact process at the critical point, where all temporal quantities have to scale in the same way, the existence of such a crossover time is impossible, meaning that the two limits have to give identical results. In this sense critical systems are atypical examples of ageing phenomena in so far as the crossover from `equilibrium' to ageing is part of the scaling function $f_R(\xi)$, while in ordinary ageing phenomena it is not.

In the case of the critical contact process the equivalence of the two limits implies that for $\xi\to 1$ the scaling function $f_R(\xi)$ has to reproduce the behavior of the quasi-stationary response function. Since in this limit one measures the response to a local activation in the absorbing state, the response function $R(t,s)$ tends to the so-called pair-connectedness function $c(t,s)$ (see e.g.~\cite{Hin}) measured at the same positions in space. In critical DP this function is known to decay algebraically with the time difference as $c(t,s)\sim (t-s)^{-2 \delta}$, fixing the exponent in the second line of Eq.~(\ref{AsymptoticPowerLaw}). 

The main point of Ref.~\cite{Malte1}, on which we will focus in the present work, is to apply a generalization of ordinary dynamical scaling to a space-time dependent kind of scale invariance, known as the theory of \textit{local scale invariance} (LSI)~\cite{Malte0}. The central assumption of this theory is that the linear response function is given by correlation functions of so-called quasi-primary fields which transform covariantly under the action of a group of local scale transformations. As in the case of conformal invariance, this extended  symmetry fully determines the functional form of certain universal scaling functions. In particular, it is predicted that the autoresponse function is given by the explicit expression~\cite{Malte0,Ising1}
\begin{equation}
\label{ExplicitForm}
R(t,s)\;=\;r_0 \, \left( \frac{t}{s} \right)^{1+a -\lambda_R/z} \, (t-s)^{-1-a}\,,
\end{equation}
where $r_0$ is a normalization constant. This result implies that the scaling function $f_R(\xi)$ has to be the form
\begin{equation}
\label{LSI}
f_R(\xi) \;\propto\; \xi^{-\lambda_R/z} \, (1-1/\xi)^{-1-a}\,,
\end{equation}
which LSI predicts to hold \textit{exactly} for all $\xi>1$. Note that the exponents $a$ and $\lambda_R/z$ are \textit{not} predicted by the theory, they rather depend on the specific model under consideration. 

Comparing this prediction with the numerical estimates of $f_R(\xi)$ for the contact process Enns \textit{et al.} observed a ``perfect agreement almost down to $t/s=1$'' (see Fig. 8 of Ref.~\cite{Malte1}), concluding that LSI is a symmetry of the critical contact process. In a subsequent paper, Ramasco \textit{et al.} confirmed these findings by Monte-Carlo simulations~\cite{Malte2}. Moreover, they were able to prove Eq.~(\ref{ExplicitForm}) within a mean field approximation of DP, which is expected to be valid in space dimensions $d \geq 4$. 

However, there are various reasons to cast doubt on these results. In dimensions $d<4$ the phase transition of DP is described by an interacting field theory with non-trivial loop corrections~\cite{Tauber}. To establish LSI as an \textit{exact} symmetry of DP (analogous to conformal invariance in equilibrium models) would be highly non-trivial since the field theory would have to be invariant under local scale transformations to \textit{all} orders of perturbation theory. A similar situation is encountered  in the two-dimensional kinetic Ising model with heat bath dynamics quenched from infinite to critical temperature. While Monte-Carlo simulations~\cite{Ising1,Ising2} seemed to be in agreement with the predictions of LSI, Calabrese and Gambassi~\cite{Calabrese1,Calabrese2} were able to identify higher-order contributions of the corresponding field theory that violate LSI. More recently the existence of corrections to LSI in the Ising model could be confirmed numerically~\cite{Pleimling}. These findings suggest to search for a similar failure of this theory in the case of DP.

Regarding possible violations of local scale invariance, the most sensitive part of the predicted scaling function~(\ref{ExplicitForm}) is the knee of the crossover from one power law to the other, which takes place at $t/s \approx 2$. We therefore present a direct measurement of $R(t,s)$ in the 1+1-dimensional critical contact process, restricting the simulations to the range $1 < t/s < 4$. As usual in numerical simulations, scaling can be observed only when $s$, $t$, and $t-s$ are much larger than the microscopic reference time scale. Clearly, such deviations are most pronounced close to $t/s=1$ when $t-s$ is small. In order to control such errors we generate different data sets, increasing $s$ by powers of~$2$. Collapsing these data sets allows us to get an impression of the range from where on scaling sets in.

\begin{figure}
\centerline{\includegraphics[width=140mm]{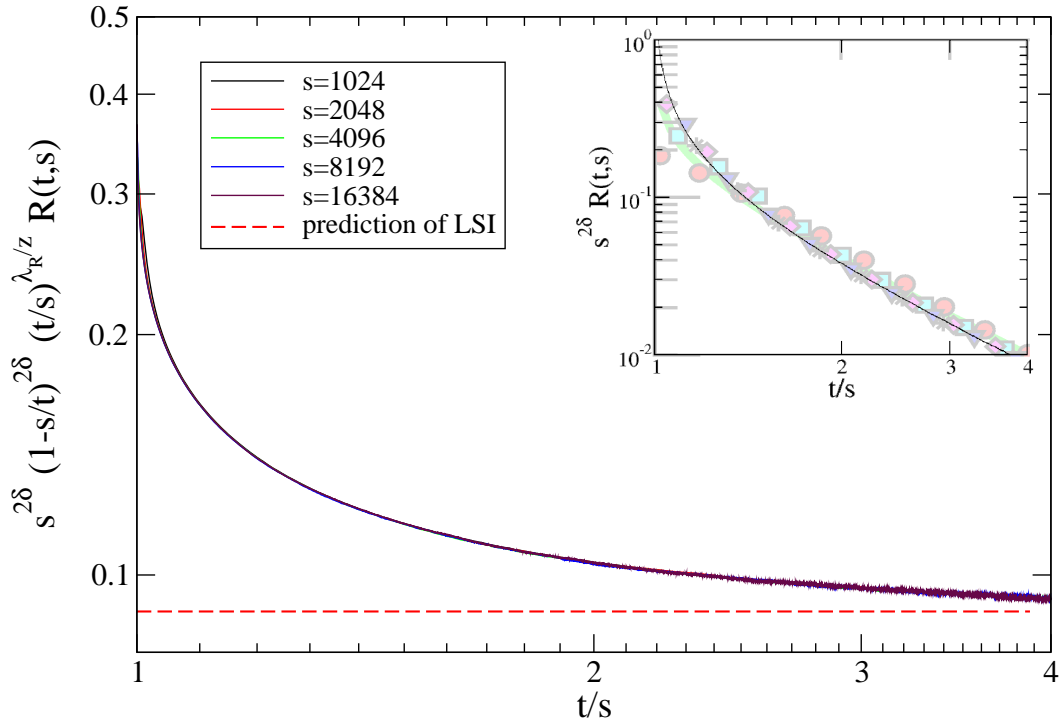}}
\caption{\label{FIGSCALED} \small
Data collapse of the autoresponse function $R(t,s)$ in a 1+1-dimensional contact process at criticality for $s=1024$, $2048, \ldots, 16384$, using the DP exponent $\delta\simeq 0.15946$. The numerical results differ significantly from the theoretical prediction of local scale invariance according to Eq.~(\ref{LSI}), which is shown as a dashed line. In order to illustrate how the data complies with earlier results, the inset shows the present data on top of a zoomed part of Fig.~8 in Ref.~\cite{Malte1}.
}
\end{figure}

These collapsed data sets (generated by a C++ program added as an attachment) are shown in Fig. 1. In order to visualize deviations from the prediction of LSI, we plotted $R(t,s) \, s^{2 \delta} (1-\frac{1}{t/s})^{2 \delta}\,(t/s)^{\lambda_R/z}$ versus $t/s$ in a double-logarithmic plot. According to Eq.~(\ref{LSI}), LSI predicts this quantity to be constant, as indicated by the horizontal dashed line. However, the data collapse clearly demonstrates that there is no such coincidence. As can be seen, these deviations are robust and persist as the numerical effort is enhanced by increasing $s$. This means that the predicted scaling form~(\ref{LSI}) does not hold in the case of directed percolation, disproving the conjectures of Refs.~\cite{Malte1} and \cite{Malte2}.

More recently, Picone, Henkel, and Pleimling proposed a \textit{generalized} scaling form~\cite{Malte3,Malte4} derived from the generators of the Schr{\"o}dinger algebra. It was argued that for applications to ageing, invariance under time translations is actually not needed, instead it suffices to require covariance of the response function under the subalgebra without time-translations. This assumption then leads to a generalized scaling form 
\begin{equation}
\label{GLSI}
f_R(\xi) \;\propto\; \xi^{-\lambda_R/z} \, (1-1/\xi)^{-1-a'}\,,
\end{equation}
with an \textit{independent} exponent $a'$ different from $a$. This generalized scaling form was successfully applied to the one-dimensional Glauber Ising model at zero temperature~\cite{Glauber1,Glauber2,Glauber3,Malte3}, the OJK approximation~\cite{Berthier,Mazenko,Malte4}, and to the critical Ising spin glass in three dimensions~\cite{Malte4}. However, in the case of the critical contact process this scaling form is inconsistent since for $a' \neq a$ it does not reproduce the correct asymptotic behavior of Eq.~(\ref{AsymptoticPowerLaw}) in the limit $\xi\to 1$.
 
\begin{figure}
\centerline{\includegraphics[width=160mm]{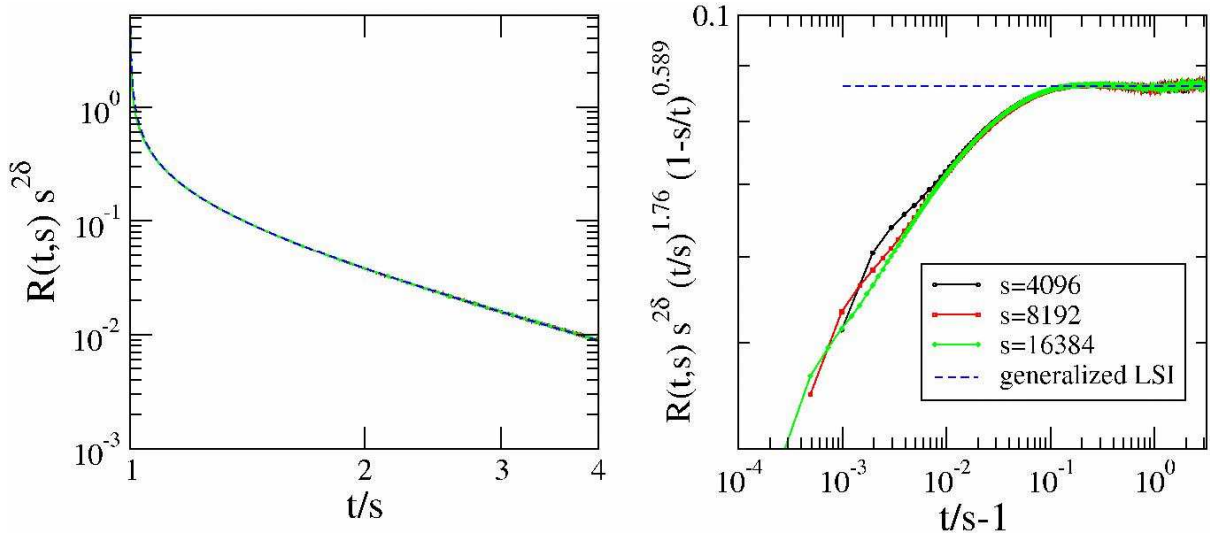}}
\caption{\label{FIGSCALED} \small
Left: Fit of the generalized scaling function in Eq.~(\ref{GLSI}) on top of the collapsed numerical data in the same representation as in Fig. 1. As can be seen, there is an excellent coincidence over a wide range of $t/s$. Right: The same data sets plotted against $t/s-1$ scaled in such a way that the prediction according to Eq.~(\ref{GLSI}) would lead to a horizontal line. The sharp decline for $t/s \to 1$ illustrates why this scaling form is incompatible with the quasi-stationary limit in DP (see text). This decline is stable as $s$ is increased so that scaling corrections due to a microscopic reference time as its origin can be ruled out.
}
\end{figure}

This inconsistency can also be seen numerically. Using the exponents $\lambda_R/z$ and $a'$ as fit parameters\footnote{Note that $a$ cannot be used as a fit parameter because deviations from the DP value $1+a=2\delta=\beta/\nu_\parallel$ would immediately destroy collapse of curves for different $s$.} it is possible to find an excellent coincidence with the numerical data for $a'\approx0.41$ and $\lambda_R/z\approx 1.76$, as demonstrated in the left panel of Fig. 2. However, according to the previous argument such a fit has to fail in the limit $\xi \to 1$. As demonstrated in the right panel of Fig. 2, this is indeed the case, irrespective of scaling corrections due to a microscopic reference time. Similar deviations can be observed in LCTMRG calculations~\cite{Enss}. 

The failure of the generalized scaling form~(\ref{GLSI}) with $a \neq a'$ in the contact process may be surprising since in various other models it seems to work very well. For example, the generalized scaling form was found to hold exactly in the case of the one-dimensional Glauber-Ising at zero temperature (and equivalently the annihilation process), which is also a scale-free process so that the aforementioned limits are (a) and (b) are identical~\cite{Zannetti,Mayer}. However, in the one-dimensional Ising model at $T=0$ the magnetic field can only flip spins next to a domain wall whose density is known to decay as $s^{-1/2}$. Therefore, in the limit $s \to \infty$ with $t-s$ fixed the response function goes to zero as $R(t,s) \sim s^{-1/2} (t-s)^{-1/2}$. Taking the factor $s^{-1/2}$ into account the generalized scaling form can be applied consistently in this case. Contrarily, in the contact process the external field can create activity at any inactive site in the bulk so that the response function tends to $R(t,s) \sim (t-s)^{-2\delta}$ without an $s$-dependent prefactor, implying $a=a'$. Similarly, in the 2d or 3d Ising model at criticality, the magnetic field can flip any spin in the bulk, hence a numerical fit of the generalized scaling function with $a \neq a'$ is expected to be inconsistent in the limit $\xi \to 1$.

\vglue 2mm

To summarize, we have shown that the original prediction of LSI in Eq.~(\ref{LSI}) fails in the case of the 1+1-dimensional contact process at criticality. In so far the contact process is akin to the two-dimensional Ising model at the critical point, where similar discrepancies where found explicitly~\cite{Calabrese1,Calabrese2,Pleimling}. A recently proposed \textit{generalized} scaling form based on LSI without translational invariance produces surprisingly accurate results over a wide range of $t/s$ if the exponents are fitted numerically. However, this generalized scaling form is inconsistent with the known asymptotic behavior of DP, leading to systematic deviations in the limit $t/s\to 1$ which cannot be eliminated by increasing the numerical effort. These results strongly suggest that LSI in the present form (interpreted as a symmetry analogous to conformal invariance in equilibrium which determines the scaling function entirely) is \textit{not} an exact symmetry of 1+1-dimensional DP at criticality. 
 
As already conjectured in the context of the 2d Ising model~\cite{Calabrese2,Pleimling}, these  discrepancies indicate that LSI is not a generic property of ageing phenomena, rather this theory seems to be restricted to systems with a Gaussian response, including diffusive models ($z=2$), off-critical processes, and the mean-field limit (tree level) of certain non-equilibrium critical phenomena above the upper critical dimension.


\vglue 8mm

{\bf \noindent Note:}\\
After submission, Enns, Henkel, and Pleimling released a preprint [cond-mat/0605211v1] in which they interpret the deviations shown in the right panel of Fig.~2 as a crossover \textit{within the scaling function itself} that takes place at some value of the \textit{quotient} $\xi_c=t/s$, proposing that LSI should hold exactly above $\xi_c$. 

\vglue 2mm


\vglue 2mm

{\bf \noindent Acknowledgments:}\\
I would like to thank A. Gambassi, M. Henkel, S. L{\"u}beck and M. Pleimling for interesting discussions and helpful comments. Special thanks go to T. Enss, who identified a bug in the first version of the program code.

\vglue 2mm


\end{document}